\documentclass{aa}
\usepackage[pdfpagelabels=false]{hyperref}
\hypersetup{colorlinks=true,
            linkcolor=blue,
            citecolor=blue,
            filecolor=blue,
            urlcolor=blue}
\usepackage{graphicx}
\usepackage{bm}
\usepackage{pdflscape}
\usepackage[T1]{fontenc}
\usepackage{ae,aecompl}
\usepackage{newtxtext,newtxmath}
\usepackage{orcidlink}
\usepackage{framed}
\usepackage{comment}
\usepackage{cancel}
\usepackage[normalem]{ulem}
\usepackage{enumerate}
\usepackage{natbib}
\bibpunct{(}{)}{;}{a}{}{,}
\DeclareRobustCommand{\r}[1]{{\rm #1}}
\DeclareRobustCommand{\Sec}[1]{Sect.~\ref{#1}}
\DeclareRobustCommand{\Tab}[1]{Table~\ref{#1}}
\DeclareRobustCommand{\Fig}[1]{Fig.~\ref{#1}}
\DeclareRobustCommand{\Eq}[1]{Eq.~(\ref{#1})}

\newcommand{\be}{\begin{equation}}
\newcommand{\ee}{\end{equation}}
\newcommand{\bea}{\begin{eqnarray}}
\newcommand{\eea}{\end{eqnarray}}

\usepackage{xcolor}
\definecolor{darkgreen}{rgb}{0.0, 0.7, 0.0}

\newcommand{\rd}{{\rm d}}

\begin{document} 

\title{Sensitivity of the Hongmeng 21cm experiment to scattering dark matter}

\author{
	Junsong Cang \inst{1,2} \thanks{cangjunsong@outlook.com}
	\orcidlink{0000-0002-0061-0728}
	\and
	Yu Gao \inst{3}
	\orcidlink{0000-0002-8228-9981}
	\and
	Yin-Zhe Ma \inst{4,5}
	\orcidlink{0000-0001-8108-0986}
}

\institute{
    Theoretical and Scientific Data Science, Scuola Internazionale Superiore di Studi Avanzati (SISSA), Via Bonomea 265, 34136 Trieste, Italy
    \and
    Scuola Normale Superiore, Piazza dei Cavalieri 7, 56126 Pisa, Italy
    \and
    State Key Laboratory of Particle Astrophysics, Institute of High Energy Physics, Chinese Academy of Sciences, Beijing, 100049, China
    \and
    Department of Physics, Stellenbosch University, Matieland 7602, South Africa
    \and
    National Institute for Theoretical and Computational Sciences (NITheCS), Stellenbosch University, Matieland 7602, South Africa
}
    
\date{Received XXXX; accepted XXXX}

\abstract{
Scattering between dark matter and baryons can cool the intergalactic medium temperature and deepen the 21cm signal. Such interactions have been proposed to explain the unusually deep 21cm absorption signal reported by EDGES in 2018. We explore the potential to detect dark matter - baryon scattering with the Hongmeng project, an upcoming Moon-orbiting satellite experiment dedicated to measuring the global 21cm signal between redshifts $11-46$. We self-consistently forward-model the simulated sky-temperature data, jointly varying the astrophysical and foreground models. We show that even with a very conservative observational strategy in which the experiment only takes data when the Earth and the Sun are both shielded by the Moon, Hongmeng can tighten the current constraints on the cross section of dark matter - baryon scattering $\sigma_0$ by a factor of 39 after the full mission, which lasts for five years. The prospective upper limit on $\sigma_0$ can reach $5.4 \times 10^{-43} {\rm cm^2}$ for dark matter masses between 0.1 MeV and 0.3 GeV. Even after only one month of operation, an improvement by a factor of 4 relative to current $\sigma_0$ limits can be expected.
}

\maketitle

\section{Introduction}
The redshifted 21cm signal holds the potential to revolutionize cosmology with the promises of mapping out the first half of our observable Universe.
The first claim of a detected 21cm signal was reported by the Experiment to Detect the Global Epoch of Reionization Signature (EDGES) in 2018 \citep{Bowman:2018yin},
featuring a global (i.e., spatially averaged) absorption signal centered at redshift 17 with a depth of -500 mK,
approximately twice as deep as the most optimistic astrophysical prediction in which the cosmic microwave background (CMB) dominates the radio background and the intergalactic gas cools adiabatically.
Such a deep signal can be achieved by either a radio background in excess of the CMB \citep{Feng:2018rje, Ewall-Wice:2018bzf, Fialkov:2019vnb, Mirocha:2018cih, Ewall-Wice:2019may, Reis:2020arr, Cang:2024gtp},
or excess gas cooling via scattering between dark matter (DM) and baryons \citep{Munoz:2018pzp, Barkana:2018lgd}.
Early dark energy can also achieve the additional cooling needed to reproduce the EDGES depth,
but these models are strongly ruled out by other cosmological observations \citep{Hill:2018lfx},
such as the CMB anisotropy spectrum \citep{Karwal:2016vyq} or
measurements of the Hubble constant \citep{Riess:2018uxu}.

Dark matter decouples much earlier than baryons owing to its extremely weak or even entirely absent interactions with standard model particles, 
and its temperature therefore remains much lower than that of baryons throughout cosmic history.
In a scattering dark matter (SDM) framework,
DM can interact with baryons through elastic scattering. This allows some heat transfer from baryons to DM, 
leading to excess cooling of the gas, and consequently, to a deepened 21cm absorption signal that can potentially reproduce the depth reported by EDGES, provided that a strong Lyman-$\alpha$ background was present to couple the gas spin temperature to its kinetic temperature \citep{Munoz:2018pzp, Barkana:2018lgd}.
In addition to deepening the global signal,
cooling from the SDM can also enhance the 21cm signal fluctuation \citep{Munoz:2018jwq},
offering promising prospects for experiments dedicated to measuring the 21cm signal power spectrum \citep{Rahimieh:2025lbf},
such as the Square Kilometre Array (SKA; e.g., \citealt{Mellema:2012ht}), the Hydrogen Epoch of Reionization Array (HERA; \citealt{HERA:2021noe}), 
or the Low Frequency Array (LOFAR; \citealt{LOFAR:2013jil}).
Models made to achieve such interactions generally fall into two categories \citep{Flitter:2023mjj}:
(i) millicharged models \citep{McDermott:2010pa,Munoz:2018pzp,Kovetz:2018zan, Berlin:2018sjs, Barkana:2018qrx, Slatyer:2018aqg, Liu:2018uzy, Falkowski:2018qdj, Munoz:2018jwq,Driskell:2022pax,Zhang:2024mmg},
which require the SDM to carry a fractional electrical charge and only interact with charged particles, that is, free electrons and protons; and
(ii) baryophilic models \citep{Dvorkin:2013cea, Munoz:2015bca, Boddy:2018wzy, Fialkov:2018xre, Xu:2018efh, Short:2022bmm, He:2023dbn},
in which case the SDM interacts with all standard model particles.

This shows that the 21cm signal provides a complementary probe of the SDM in addition to underground direct-detection experiments such as PandaX \citep{PandaX-4T:2021bab,PandaX:2024qfu},
China Dark Matter Experiment (CDEX) \citep{CDEX:2018lau,CDEX:2019hzn,CDEX:2022kcd}, and LUX-ZEPLIN (LZ) \citep{LZ:2022lsv,LZ:2024zvo},
which detect the SDM by measuring nuclear recoils caused by collisions with DM.
The SDM can also be constrained by a range of cosmological probes \citep{Jaeckel:2010ni,Ali-Haimoud:2021lka,Zhang:2024mmg,Lin:2023yqn} other than the 21cm signal.
For example,
DM-baryon scattering can suppress the formation of small-scale structure,
and stringent SDM limits can be derived using measurements of the matter power spectrum from the CMB and the Lyman-$\alpha$ forest,
or by abundance measurements of Milky Way satellite galaxies \citep{Dvorkin:2013cea,Gluscevic:2017ywp,Nguyen:2021cnb,Buen-Abad:2021mvc}.

Despite the excitement surrounding a possible discovery,
there are concerns about the modeling of EDGES data, however.
~\citet{Hills:2018vyr} pointed out that the parameters of foreground model that EDGES used to model their data have unphysical values.
The phenomenological Flat-Gaussian 21cm signal model applied in the original analysis was highly unlikely for any realistic astrophysical models without resorting to extreme fine-tuning \citep{Mittal:2022twx,Gessey-Jones:2023amq,Nelander:2025tda}.
\citet{Sims:2019kro} and \citet{Cang:2024gtp} suggested that there could be calibration residuals in EDGES sky temperature data and that after consistently forward modeling the data with physically motivated models for cosmic 21cm signal,
foreground and calibration,
the inferred 21cm signal was in fact entirely consistent with that expected from standard model.
Furthermore,
a subsequent measurement made by the Shaped Antenna measurement of the background RAdio Spectrum (SARAS) \citep{Singh:2021mxo, Bevins:2022ajf} pointed to a nondetection of a cosmic 21cm signal.

A robust detection of the 21 cm signal remains extremely challenging due to the complex interplay of astrophysical and terrestrial foregrounds, 
which are typically brighter by some orders of magnitude than the cosmological signal itself \citep{Pritchard:2011xb,Hibbard:2023sdv,Mittal:2024mzv}. 
Reliable measurements therefore require a careful joint characterization of the cosmic signal, foregrounds, and instrumental effects, along with validations from independent observations.
Ground-based global signal experiments such as EDGES \citep{Bowman:2018yin}, 
SARAS \citep{Singh:2021mxo, Bevins:2022ajf}, 
and Large Aperture Experiment to Detect the Dark Ages (LEDA; \citealt{Greenhill:2012mn}) are susceptible to  various terrestrial contaminations \citep{Shi:2022zdx},
for instance, refraction and attenuation of the radio wave by the Earth ionosphere \citep{Vedantham:2013mea, Shen:2020jwt, 2023ChJSS..43...43C},
antenna interactions with the nearby environment \citep{Shi:2022zdx},
and radio frequency interference (RFI) \citep{Tauscher:2020wso},
all of which make a precise measurement of the 21cm signal even more difficult.
Several space-borne missions have been proposed to circumvent these issues,
for instance, the Dark Ages Polarimetry PathfindER (DAPPER; \citealt{Tauscher:2018uxi}),
the Dark Ages Radio Explorer (DARE; \citealt{Burns:2017ndd}),
the Farside Array for Radio Science Investigations of the Dark ages and Exoplanets (FARSIDE; \citealt{Burns:2021pkx})
and Hongmeng (also known as Discovering the Sky at the Longest wavelength, or DSL) \citep{Chen:2019xvd, Chen:2020lok, Shi:2022zdx, 2023ChJSS..43...43C}.

Designed to operate on a lunar orbit,
the Hongmeng mission \citep{Chen:2019xvd, Chen:2020lok, Shi:2022zdx,2023ChJSS..43...43C} is naturally free from ionosphere contamination,
and most importantly, the Moon can serve as a natural shield against RFI contamination from the Earth,
which remains a major challenge even for Earth-orbiting satellites.
Hongmeng is designed to measure the 21cm signal in a broad redshift window of $11 \lesssim z \lesssim 46$ \citep{Chen:2020lok},
reaching deep into the cosmic dark ages while encompassing the onset of reionization.
New physics processes such as DM–baryon scattering might have operated during the dark ages, 
well before the onset of astrophysical heating.
The ability to explore this pristine epoch places Hongmeng in a particularly advantageous position to provide a clean probe of new physics, 
free from astrophysical uncertainties,
and \citet{Zhao:2024jad} showed that with a modest observing time, Hongmeng might constrain DM annihilation and decay, as well as Hawking radiation from primordial black holes, with a precision exceeding current leading limits by several orders of magnitude.

We investigate the potential of Hongmeng in probing DM-baryon scattering.
We consistently forward-model the cosmic signal and foreground.
The rest of this paper is organized as follows:
we briefly review the 21cm signal and impact of SDM in \Sec{Sec_21cm_and_SDM},
our forecast pipeline is detailed in \Sec{Sec_Inference},
we show our results in \Sec{Sec_Results}, and we finally conclude in \Sec{Sec_Discussions}.
We use a standard $\Lambda$CDM cosmology throughout with the relevant cosmological parameters set by the Planck 2018 results \citep{Planck:2018vyg}:
$h = 0.6766$,
$\Omega_{\rm b} h^2 = 0.02242$,
$\Omega_{\rm c} h^2 = 0.11933$,
$\ln (10^{10}A_{\rm s}) = 3.047$, and
$n_{\rm s} = 0.9665$.

\section{21cm signal with scattering dark matter}
\label{Sec_21cm_and_SDM}
In the cosmological context,
the strength of 21cm signal is measured by its differential brightness temperature $T_{21}$ (referred to as 21cm temperature hereafter for simplicity).
Ignoring the spatial fluctuations,
global $T_{21}$ can be approximated as \citep{Furlanetto:2004zw,Mesinger:2010ne,Pritchard:2011xb,Sitwell2014},
\be
T_{21}
\simeq
27 
x_{\rm H}
\left(
1 - 
\frac{T_{\rm R}}{T_{\rm s}}
\right)
\left(
\frac{1+z}{10}
\frac{0.15}
{\Omega_{\rm m}h^2}
\right)^{1/2}
\left(
\frac{\Omega_{\rm b}h^2}
{0.023}
\right)
{\rm mK},
\label{Eq_T21}
\ee
where $x_{\rm H}$ is the hydrogen neutral fraction,
$T_{\rm R}$ is the radio background temperature and was assumed to be dominated by the CMB, such that $T_{\rm R} = T_{\rm CMB}$, and $T_{\rm CMB} = 2.728 (1+z)\ {\rm K}$ is the CMB temperature.
$\Omega_{\rm m}$ and $\Omega_{\rm b}$ are the fractional energy density in matter and baryons, respectively,
$h$ is the Hubble constant in units of $100 {\rm km\ s^{-1} Mpc^{-1}}$,
and finally, the spin temperature $T_{\rm s}$ is coupled to the baryonic intergalactic medium (IGM) kinetic temperature $T_{\rm k}$ and the radio background temperature $T_{\rm R}$ by \citep{Pritchard:2011xb,Sitwell2014},
\be
T^{-1}_{\rm s}
=
\frac{
T^{-1}_{\rm R}
+
x_\alpha
T^{-1}_\alpha
+
x_{\rm c}
T^{-1}_{\rm k}
}
{1 + x_\alpha + x_{\rm c}},
\label{Eq_Ts}
\ee
where $T_\alpha \approx T_{\rm k}$ is the color temperature \citep{Hirata:2005mz},
and $x_{\rm c}$ and $x_\alpha$ are the collisional and Wouthuysen-Field coupling coefficients, respectively (see \citealt{Pritchard:2011xb}).

We used the publicly available {\tt 21cmFirstCLASS} code \citep{Flitter:2023mjj} to compute the 21cm signal.
{\tt 21cmFirstCLASS} uses a modified version of the code {\tt CLASS} \citep{Blas:2011rf, Gluscevic:2017ywp, Boddy:2018kfv,Nguyen:2021cnb,Boddy:2022tyt} to compute the initial conditions in the presence of SDM.
These are then passed to a modified version of {\tt 21cmFAST} \citep{Mesinger:2010ne, Park:2018ljd, Murray:2020trn},
which is highly customized to incorporate SDM cooling while maintaining the original astrophysical framework,
which solves for various components required to calculate $T_{21}$.
Below, we briefly review the essential ingredients related to SDM cooling in {\tt 21cmFirstCLASS} and the astrophysical background in {\tt 21cmFAST},
and we refer to \citet{Flitter:2023mjj} and \citet{Mesinger:2010ne, Murray:2020trn, Qin:2020xyh,Munoz:2021psm} for the full details.

\subsection{Scattering dark matter}
\label{Sec_SDM}

We considered a Coulomb-like interaction \citep{Boddy:2018kfv,Slatyer:2018aqg,Lin:2023yqn} that is enhanced at low redshift. In this case, the DM-baryon scattering cross section $\sigma$ takes the form
\be
\sigma = \sigma_0 \left(
\frac{v}{c}
\right)^{-4},
\label{Eq_sigma_0}
\ee
where $c$ is the speed of light,
$v$ is the relative velocity between the SDM and baryons, and
$\sigma_0$ normalizes the cross section and is treated as a free parameter.
The SDM impacts the 21cm signal primarily by changing the IGM temperature $T_{\rm k}$ in \Eq{Eq_Ts}.
In the presence of the SDM,
the evolution equation of the IGM temperature can be schematically written as \citep{Flitter:2023mjj}
\be
\frac{\r{d} T_\r{k}}{\r{d} t}
=
\left.\frac{\r{d} T_\r{k}}{\r{d} t}\right|_{\rm cosmo}
+
\left.\frac{\r{d} T_\r{k}}{\r{d} t}\right|_{\rm astro}
+
\left.\frac{\r{d} T_\r{k}}{\r{d} t}\right|_{\rm SDM},
\ee
where $\left.{\r{d} T_\r{k}}/{\r{d} z}\right|_{\rm cosmo}$ contains the cosmological contributions from Compton heating
and adiabatically cooling due to cosmic expansion,
$\left.{\r{d} T_\r{k}}/{\r{d} z}\right|_{\rm astro}$ describes the X-ray heating from astrophysical sources in {\tt 21cmFAST} \citep{Mesinger:2010ne, Park:2018ljd, Murray:2020trn},
and $\left.{\r{d} T_\r{k}}/{\r{d} z}\right|_{\rm SDM}$ is the evolution rate from the SDM \citep{Flitter:2023mjj},
\be
\left.\frac{\r{d} T_\r{k}}{\r{d} t}\right|_{\rm SDM}
=
\Gamma_\chi (T_\chi - T_\r{k})
+
\frac{2}{3 k_\r{B}}
\frac{\rho_\chi v}
{\rho_\r{b} + \rho_\chi}
\sum_t
\frac{m_\chi m_\r{b}}
{m_\chi + m_\r{t}}
D_\r{t}
(v),
\label{Eq_SDM_cooling_rate}
\ee
where the subscript $\chi$ represents the SDM particle,
$m_\chi$ and $T_\chi$ are the mass and temperature of the SDM, and
$\rho_\r{b}$ and $\rho_\chi$ are the energy densities in baryons and the SDM, respectively.
For simplicity, we assumed that all DM consists of the SDM,
and thus, $\rho_\chi = \Omega_\r{c} \rho_\r{cr} (1+z)^3$,
where $\Omega_\r{c}$ is the fractional DM density and
$\rho_\r{cr}$ is the critical density today.
$k_\r{B}$ is Boltzmann’s constant, $m_{\rm b}=m_{\rm H}/\left[x_{\rm H}\left(1-(1-m_{\rm H}/m_{\rm He})Y_{\rm He} \right) \right]$ is the mean baryon mass,
where $m_\r{H}$ and $m_\r{He}$ are the masses of hydrogen and helium atoms, respectively,
$Y_\r{He} \approx 0.245$ is the helium mass fraction.
The subscript $t$ indicates the target particle with which the SDM interacts.
We assumed a baryophilic scenario \citep{Dvorkin:2013cea, Munoz:2015bca, Boddy:2018wzy, Fialkov:2018xre, Xu:2018efh, Short:2022bmm, He:2023dbn} in which the SDM interacts with all standard model particles.

Finally, in \Eq{Eq_SDM_cooling_rate}, $\Gamma_\chi$ and $D_\r{t}$ represent the energy transfer rate and the mutual drag force acting on SDM-baryon fluids, respectively \citep{Flitter:2023mjj},
\be
\Gamma_\chi
=
\sqrt{\frac{2}{\pi}}
\frac{2 \sigma_0c^4\rho_\chi}
{3 n_\r{b}}
\sum_\r{t}
\frac{\rho_\r{t} \r{exp}({-r_\r{t}^2}/2)}
{(m_\r{t} + m_\chi)^2} u^3_{\chi t},
\ee

\be
D_\r{t}
=
\frac{(\rho_\r{b} + \rho_\chi) \sigma_0 c^4}
{\rho_\r{b} u^2_{\chi t}}
\frac{\rho_\r{t} F(r_t)}
{m_\r{t} + m_\chi}
\ee
where $n_\r{b} = \rho_\r{b}/m_\r{b}$ is the baryon number density,
$r_t = v/u_{\chi \r{t}}$, where $u_{\chi t}$ is the thermal velocity,
$
F(r_\r{t})
=
r^{-2}_\r{t}
\left[
\r{erf}
\left(
\frac{r_\r{t}}
{\sqrt{2}}
\right)
-
\sqrt{
\frac{2}{\pi}
}
r_\r{t}
\r{e}^{-r^2_\r{t}/2}
\right]
$.

Equation~(\ref{Eq_SDM_cooling_rate}) shows that SDM can either cool or heat the IGM.
When the SDM temperature is much lower than that of the baryons, the energy transfer between them acts as a coolant, lowering the IGM temperature.
However, if there exists a non-negligible relative velocity between the SDM and baryons, 
the frictional dissipation between the two fluids can heat both components.
In some cases, the SDM might therefore heat the baryons if the frictional heating is strong enough \citep{Munoz:2015bca}.
We discuss this further in \Sec{Sec_Results}.

\subsection{Astrophysical background}
\label{Sec_Astro_Background}

We assumed that stars form in population II (Pop II) and population III (Pop III) galaxies.
In {\tt 21cmFAST},
these galaxies are classified depending on their gas accretion mechanisms\citep{Qin:2020xyh, Munoz:2021psm}.
The metal-poor Pop III galaxies accrete gas mainly by rotational-vibrational transitions of hydrogen molecules,
whereas Pop II galaxies are typically made from the metallic remnants of Pop III galaxies and obtain gas primarily through the line transitions of hydrogen atoms.

Astrophysical sources affect the 21cm signal mainly through three radiative processes:
(i) IGM heating by X-ray;
(ii) ionization by UV photons; and
(iii) Lyman-$\alpha$ radiation, which couples $T_{\rm s}$ to $T_\r{k}$ via Wouthuysen-Field (WF) effects.
All these radiation fields trace the star formation history \citep{Fragos:2012vf,Park:2018ljd}.
To account for these uncertainties,
we adopted the astrophysical framework detailed in ~\cite{Munoz:2021psm} and varied the following {\tt 21cmFAST} astrophysical parameters
(in addition to the SDM parameters $m_\chi$ and $\sigma_0$):
\be
\{
 f_{\star, 10},\ \alpha_{\star,\r{II}},\  
f_{\star, 7},\ \alpha_{\star,\r{III}},\  
\mathcal{L}_\r{X, II},
\mathcal{L}_\r{X, III},
\ f_\r{esc, 10},
\ f_\r{esc, 7}.
\}
\ee
Together, these parameters describe the star formation, X-ray, and UV radiations of Pop II and Pop III galaxies.
We briefly review these parameters below and refer to \citet{Qin:2020xyh,Munoz:2021psm} for extensive discussions.

\subsubsection{Star formation}

In {\tt 21cmFAST},
the stellar fraction $f_\star$,
which describes the fraction of the halo baryonic mass converted into stars, is modeled in Pop II and Pop III galaxies as
\be
f_{\star, \r{II}} = 
\r{min}
\left[
f_{\star, 10}
\left(
\frac{M_\r{h}}{10^{10}M_\odot}
\right)^{\alpha_{\star \r{II}}},
1
\right],
\label{Eq_fstar_10}
\ee
\be
f_{\star, \r{III}} = 
\r{min}
\left[
f_{\star, 7}
\left(
\frac{M_\r{h}}{10^{7}M_\odot}
\right)^{\alpha_{\star \r{III}}},
1
\right],
\label{Eq_fstar_7}
\ee
where $M_\r{h}$ is DM halo mass, and
the subscripts II and III denote Pop II and Pop III galaxies respectively.
$f_{\star, 10}$,
$f_{\star, 7}$,
$\alpha_{\star \r{II}}$,
and  $\alpha_{\star \r{III}}$
are free parameters describing the normalization and power index of the stellar fraction.
Because X-ray, Lyman-$\alpha$, and UV radiation fields are all sourced by star formation, 
a higher star formation rate (SFR) accelerates the WF coupling, 
X-ray heating, and reionization. 
As a result, 
when the SFR is increased, the 21 cm signal becomes deeper prior to the X-ray heating era 
(due to stronger WF coupling that ties $T_{\rm s}$ to the cold IGM), 
but becomes shallower afterward due to enhanced X-ray heating and UV ionization. 

\subsubsection{X-ray radiation}

X-rays from galaxies are mainly emitted by high-mass X-ray binaries (HMXBs) \cite{Fragos:2012vf}.
Because these systems have relatively short lifetimes,
their comoving X-ray emissivity $\epsilon_\r{X}$ traces the star formation rate density (SFRD) \citep{Cang:2024gtp},
\be
\epsilon_\r{X}
=
L_\r{X}/\r{SFR} \times \r{SFRD},
\ee
where $L_\r{X}/\r{SFR}$ denotes the monochromatic X-ray luminosity per SFR and was assumed to follow a power-law energy spectrum.
Normalizations of $L_\r{X}/\r{SFR}$ below 2 keV are treated as a free parameter $\mathcal{L}_\r{X}$ in {\tt 21cmFAST},
\be
\mathcal{L}_\r{X, II}
\equiv
\int_{E_0}^{2 \r{keV}} \rd E \ 
L_\r{X,II}/\r{SFR},
\label{Eq_LX}
\ee
\be
\mathcal{L}_\r{X, III}
\equiv
\int_{E_0}^{2 \r{keV}} \rd E \ 
L_\r{X,III}/\r{SFR},
\label{Eq_LX3}
\ee
where $E_0 = 0.5$ keV is the threshold energy below which photons are absorbed within the host galaxy \citep{Das:2017fys}.
Following \citet{Park:2018ljd},
we adopted fiducial values of $\mathcal{L}_\r{X,II} =\mathcal{L}_\r{X,III}= 10^{40.5}\r{erg s^{-1} M^{-1}_\odot \r{yr}}$,
which is consistent with HMXB simulations in metal-poor environments \citep{Fragos:2013bfa}.

X-rays affect the 21cm signal primarily through heating,
and increasing $\mathcal{L}_\mathrm{X, II}$ and $\mathcal{L}_\mathrm{X, III}$ raises the kinetic and spin temperatures ($T_\r{k}$ and $T_\r{s}$), 
shifting the 21 cm signal from absorption to emission and moving the absorption trough to higher redshifts while enhancing the emission amplitude.
A fraction of the X-ray energy that is not deposited as heat can contribute to ionization and to the Lyman-$\alpha$ background \citep{Qin:2020xyh}, 
thereby accelerating reionization and WF coupling.

\subsubsection{Ionizing UV background}
The epoch of reionization is thought to be mainly driven by UV photons \citep{Furlanetto:2006jb}.
Similar to the stellar fraction in Eqs. (\ref{Eq_fstar_10}) and (\ref{Eq_fstar_7}), 
the escape fraction $f_\mathrm{esc}$, that is, the fraction of ionizing UV photons that escape from their host galaxies and contribute to the reionization of the IGM,
is modeled in {\tt 21cmFAST} as
\be
f_\r{esc, II} = 
\r{min}
\left[
f_{\r{esc}, 10}
\left(
\frac{M_\r{h}}{10^{10}M_\odot}
\right)^{\alpha_\r{esc}},
1
\right],
\label{Eq_fesc_10}
\ee
\be
f_\r{esc, III} = 
\r{min}
\left[
f_{\r{esc}, 7}
\left(
\frac{M_\r{h}}{10^{7}M_\odot}
\right)^{\alpha_\r{esc}},
1
\right],
\label{Eq_fesc_7}
\ee
where $f_{\r{esc}, 10}$ and $f_{\r{esc}, 7}$ $\alpha_\r{esc}$ are free parameters,
and we fixed $\alpha_\r{esc} = -0.3$ following \citet{Munoz:2021psm}.
$f_{\mathrm{esc}}$ affects the 21cm signal primarily by regulating reionization, 
which determines the neutral fraction $x_\mathrm{H}$ in \Eq{Eq_T21}.
A higher $f_{\mathrm{esc}}$ accelerates reionization, 
thereby suppressing the 21cm signal amplitude.

\section{Forecast pipeline}
\label{Sec_Inference}

\subsection{Sky temperature and foreground models}

The sky-averaged radio data $T_\r{sky}$ (hereafter referred to as the sky temperature) measured by global 21\,cm experiments contain the combined contributions of the cosmic signal $T_{21}$ and the foreground (FG) temperature $T_{\rm FG}$ \citep{Tauscher:2020wso,Cang:2024gtp},
\be
T_\r{sky}
=
T_{21}
+
T_\r{FG}.
\label{Eq_Tsky}
\ee
We modeled the FG temperature using a log-polynomial model \citep{Murray:2022uhg, Cang:2024gtp},
\be
T_\r{FG}
=
\left(
\frac{\nu}
{75 {\rm{MHz}}}
\right)^{-\beta}
\sum_{i=0}^{N_\r{FG} - 1}
p_i
\left[
\ln
\left(
\frac{\nu}
{
{75 {\rm{MHz}}}
}
\right)
\right]^{i},
\label{Eq_TFG}
\ee
where the frequency $\nu$ is related to the redshift by $z = 1.42\r{GHz}/\nu - 1$.
$\beta$ is the spectrum power index, which we fixed to 2.53 \citep{Jester:2009dw}.
$N_\r{FG}$ is the number of FG terms, and
$p_\r{i}$ are fitting coefficients which we varied in our forecast.

In the original EDGES 2018 analysis \citep{Bowman:2018yin},
a five-term FG model was used to describe the contaminations from the Galactic synchrotron spectrum and the Earth’s ionosphere,
which led to the claim of the first detection of the global 21cm signal.
However, 
several subsequent studies \citep{Hills:2018vyr,Sims:2019kro,Cang:2024gtp} have suggested that the five-term FG model lacks the flexibility to fully capture the spectral structure present in the EDGES data.
We adopted a seven-term FG model and set $N_\r{FG} = 7$,
which was found to exhibit the highest Bayesian evidence for EDGES data \citep{Cang:2024gtp}.

\subsection{Mock data and choice of fiducials}
\label{Sec_Mock_Data_and_Fiducials}
As with \Eq{Eq_Tsky},
our mock data can be expressed as the sum of the fiducial cosmic 21cm signal $T_{21, \r{data}}$ and the FG temperature $T_\r{FG, data}$,
\be
T_\r{sky, data}
=
T_{21,\r{data}}
+
T_\r{FG, data},
\ee
We computed $T_{21,\mathrm{data}}$ by setting $\sigma_0 = 0$ and adopting the fiducial astrophysical parameters from the EOS (evolution of the 21 cm structure) simulation \citep{Munoz:2021psm}, which have been shown to be consistent with observations of galaxy UV luminosity functions (UVLFs) and the reionization history.
Following \citet{Jester:2009dw,Zhao:2024jad},
we produced our mock FG by
\be
T_\r{FG, data}
=
1697.81
\left(
\frac{\nu}
{75 \r{MHz}}
\right)^{-2.53}
\r{K},
\label{Eq_TFG_data}
\ee
which corresponds to $p_0 = 1697.81 \r{K}$ and $p_i = 0$ ($i>0$) in \Eq{Eq_TFG}.

\subsection{Specifications of the Hongmeng experiment}
The Hongmeng mission will be conducted by a ten-satellite array operating in lunar orbit \citep{Chen:2019xvd,Shi:2022zdx,2023ChJSS..43...43C}.
One of the major goals of Hongmeng is to measure the global 21cm signal across 30-120 MHz frequencies.
This corresponds to a redshift window of $z \in [10.8, 46.3]$, which we adopted throughout our subsequent data analysis.
This task will be undertaken by one of the ten satellites,
whereas the rest of the satellites focus on mapping the sky below 30 MHz, collecting data, and communication with Earth.

Following \citet{Shi:2022zdx},
we modeled the experimental noise $\sigma_\r{n}$ for Hongmeng by
\be
\sigma_\r{n}
=
\frac{T_\r{sky, data} + T_\r{rcv}}
{\sqrt{\Delta \nu f_\r{eff} t}}
, 
\label{Eq_HM_Noise_Model}
\ee
where $\Delta \nu = 0.4$ MHz is the channel bandwidth.
$t$ is the mission duration,
and we introduced an observation efficiency parameter $f_\r{eff}$ that is determined by the observational strategy.
One of the main advantages of conducting 21cm experiments on the lunar orbit is that Earth can be used as a shield against RFI contamination.
We adopted a conservative observational strategy in which the measurement of the 21cm signal is only conducted when the Earth and the Sun are both shielded by the Moon.
This gives an observation efficiency of $f_\r{eff} = 0.1$ \citep{Shi:2022zdx}.
The full Hongmeng mission will last for 5 years,
and in our analysis, we considered different mission durations to show the expected scientific outputs at different mission stages.

Finally, in \Eq{Eq_HM_Noise_Model}, the receiver noise $T_\r{rcv}$ describes the equivalent noise temperature induced by the impedance mismatch between the antenna and receiver.
For the Hongmeng instruments,
it is given by \citep{Shi:2022zdx} 
\begin{eqnarray}
T_\r{rcv}
&=& 
200\,{\rm K} \nonumber \\
& \times & \left[
 16.809
\exp
\left(
-
\frac{
\left(
(\nu/75\,{\rm MHz}) - 0.4
\right)^{1.277}
}
{
0.0808}
\right) + 1.034
\right]. \nonumber \\
\end{eqnarray}
As an additional noise term,
across the frequencies of interests, $T_\r{rcv}$ is consistently lower than our fiducial sky temperature $T_\r{sky, data}$ by roughly a factor of five.
Our overall noise is therefore dominated by $T_\r{sky, data}$.

\subsection{Fisher analysis}
We used a Gaussian likelihood $\mathcal{L}$ to model the data distribution for the Hongmeng experiment,
\be
\ln \mathcal{L}
=
-\frac{1}{2}
\sum
\left[
\frac{
(T_\r{sky, data}
-
T_\r{sky, model})^2
}
{
\sigma^2_\r{n}}
+
2 \ln \sigma_\r{n}
\right]
+
{\rm const},
\label{Eq_LogLike}
\ee
where the summation is over all frequency bands covered by Hongmeng,
$T_\r{sky, data}$ is the mock data for the measured sky temperature (see \Sec{Sec_Mock_Data_and_Fiducials}), and
$T_\r{sky, model}$ is the sky temperature computed for different 21cm and FG model parameters. We used the Fisher formalism to estimate the expected uncertainties on the SDM cross section $\sigma_0$.
Specifically, our Fisher matrix $\mathcal{F}$ can be written as
\be
\mathcal{F}_\r{ij}
=
\sum_\nu
\frac{1}
{\sigma^2_\r{n}}
\frac{\partial T_\r{sky}}
{\partial \theta_\r{i}}
\frac{\partial T_\r{sky}}
{\partial \theta_\r{j}}
\ee
where $\theta_\r{i}$ denotes the model parameters.
The prospective 68\% confidence interval (C.I.) $\sigma_\r{i}$ of the parameter $\theta_\r{i}$ can be derived from the Fisher matrix by $
\sigma_\r{i}
=
\sqrt{
(\mathcal{F}^{-1})_{\r{ii}}
}
$.

\begin{table}[t]
\caption{\centering
Parameters varied in our forecast and their fiducial values.
}
\label{Tab_varied_params}
\centering
\begin{tabular}{c|ccccc}
\hline
Parameters & \ \ Eq. & \ \ Units \ \ & \ \ Flat prior \ \ & Fiducial\\
\hline
$\sigma_0$ & \ref{Eq_sigma_0}  & $\r{cm^2}$ & linear &  0\\
$m_\chi$ &\ref{Eq_SDM_cooling_rate} & GeV & - & -\\
\hline
$f_{\star,10}$ & \ref{Eq_fstar_10} & - & $\log_{10}$ & $10^{-1.25}$\\
$f_\r{esc, 10}$ & \ref{Eq_fesc_10} & - & $\log_{10}$ & $10^{-1.35}$\\
$\mathcal{L}_\r{X,II}$ &\ref{Eq_LX} & $\r{erg\ s^{-1} M^{-1}_{\odot} yr}$ & $\log_{10}$ & $10^{40.5}$\\
$\alpha_{\star, \r{II}}$ & \ref{Eq_fstar_10} & - & linear & 0.5\\
$f_{\star, 7}$ & \ref{Eq_fstar_7} & - & $\log_{10}$ & $10^{-2.5}$\\
$f_\r{esc, 7}$ & \ref{Eq_fesc_7} & - & $\log_{10}$ & $10^{-1.35}$\\
$\mathcal{L}_\r{X,III}$ &\ref{Eq_LX3} & $\r{erg\ s^{-1} M^{-1}_{\odot} yr}$ & $\log_{10}$ & $10^{40.5}$\\
$\alpha_{\star, \r{III}}$ & \ref{Eq_fstar_7} & - & linear & 0\\
\hline
$p_0$ & \ref{Eq_TFG}  & K & linear &  $1697.81$\\
$p_{1-6}$ & \ref{Eq_TFG}  & K & linear & $0$\\
\hline
\end{tabular}
\end{table}

In total,
we varied the following 21cm and FG parameters:
\be
\{
 f_{\star, 10},\ \alpha_{\star,\r{II}},\  
f_{\star, 7},\ \alpha_{\star,\r{III}},\  
\mathcal{L}_\r{X, II},
\mathcal{L}_\r{X, III},
\ f_\r{esc, 10},
\ f_\r{esc, 7},
\sigma_0,
p_\r{FG, 0-6}.
\}
\ee
The first four parameters affects the star formation rate. The fifth and sixth parameters determine the overall strength of the X-ray radiation, and the seventh and eighth parameters affect the escape fraction of the ionizing photons. The dark matter physical parameter is the ninth parameter ($\sigma_{0}$), and the remaining seven parameters are the foreground parameters. For convenience,
we summarize all these parameters and their fiducial values in \Tab{Tab_varied_params}.
All our simulations used the same random seed and were run in a box with a volume of $(200\,\mathrm{cMpc})^3$ with a resolution of $50^3$.
Apart from the astrophysical and SDM parameters listed above,
all other parameters were fixed to their default values in {\tt 21cmFirstCLASS}.

\section{Results}
\label{Sec_Results}

\begin{figure}[t]
\begin{center}
\includegraphics[width=\textwidth/2]{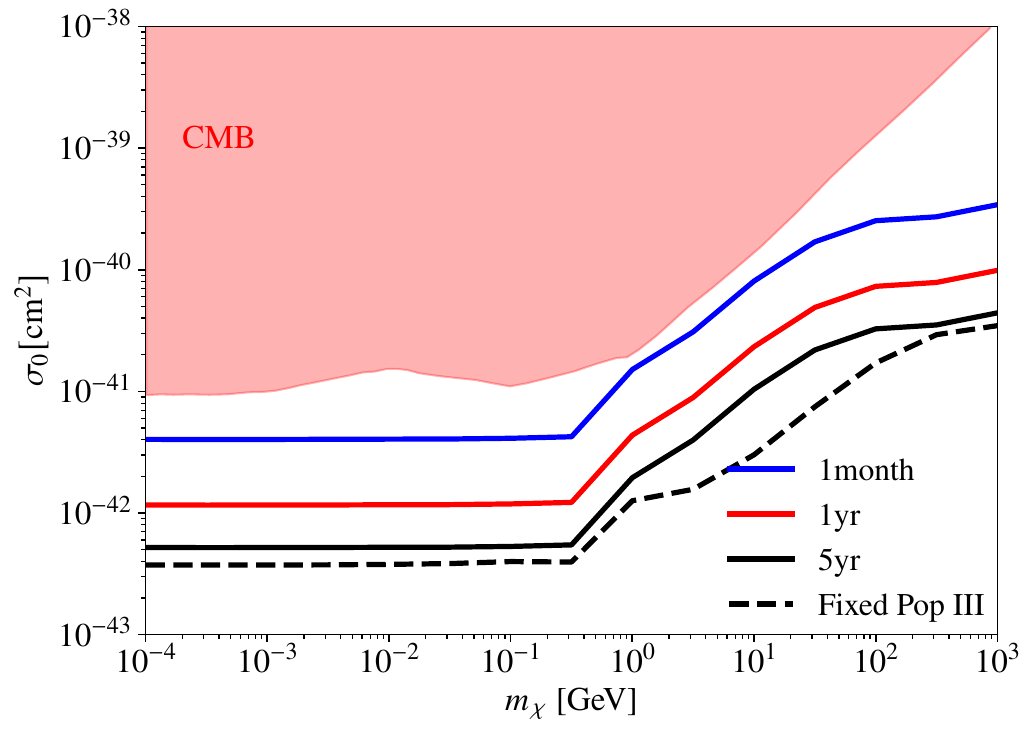}
\end{center}
\caption{
95\% C.I. upper limits on DM-baryon scattering cross section $\sigma_0$.
The solid blue and red lines assume 
mission durations of one month and one year, respectively, and
the solid black line shows the results for the full mission, which lasts for five years.
They correspond to an effective observation time of 73h, 877h, and 4383h,
respectively.
The dashed black line shows the expected five-year limits 
when Pop III parameters are fixed.
The current exclusion bound from the CMB \citep{Nguyen:2021cnb} is shown with red shaded regions.
}
\label{Fig_DM_Limits}
\end{figure}

\begin{figure*}[ht]
\centering
\includegraphics[width=\textwidth]{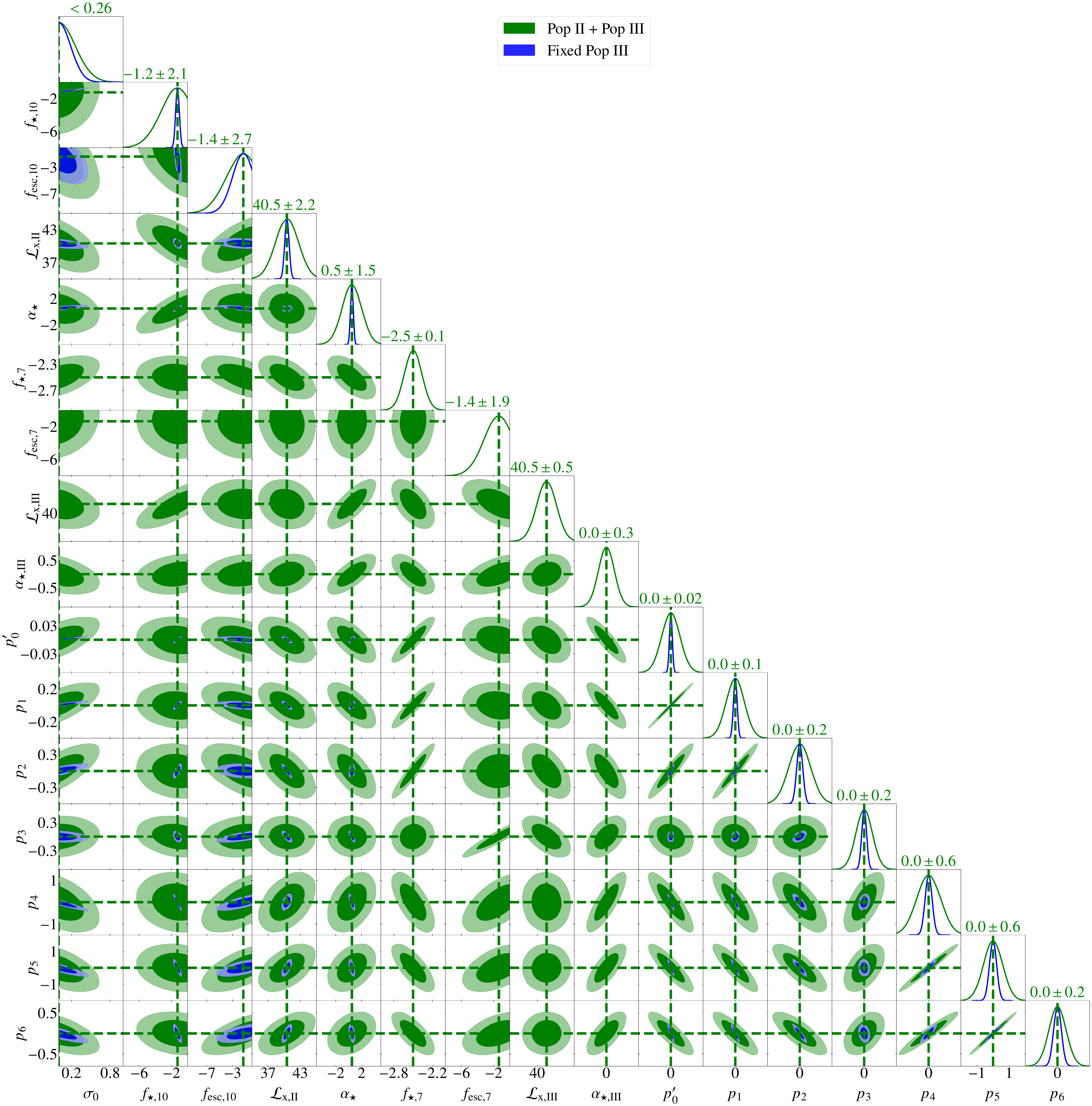}
\caption{
Prospective constraints on $\sigma_0$,
astrophysical parameters,
and FG parameters.
For compactness, $\sigma_0$ is expressed in units of $10^{-42}\r{cm^2}$.
The parameters [$f_{\star,10},\, f_{\r{esc},10},\, \mathcal{L}_{\r{X,II}},\, f_{\star,7},\, f_{\r{esc},7},\, \mathcal{L}_{\r{X,III}},\, f_{\star,7}$] are expressed in $\r{log}_{10}$ space, 
and all remaining parameters are in linear scale. 
We defined $p'_0 \equiv p_{0,\r{fid}} - p_0$, where $p_{0,\r{fid}} = 1697.81\,\r{K}$ is the fiducial value of $p_0$.
We assumed the full five-year mission duration, and $m_\chi$ was fixed to 0.1 MeV.
The dark and light shaded regions indicate the 68\% and 95\% confidence intervals,
respectively.
The green contours show the case in which we jointly varied the Pop II and Pop III parameters (Pop II + Pop III),
and the blue contours show the case in which we fixed the Pop III parameters ($f_{\star, 7}, f_\r{esc,7}, \mathcal{L}_\r{x, III}$ and $\alpha_{\star, \r{III}}$) to their fiducial values.
The constraints on Pop II, FG, and $\sigma_0$ are all significantly tightened when the Pop III parameters are fixed.
The dashed green lines show the input fiducial values for each parameter,
and the titles show the 68\% C.I. limits from the Pop II + Pop III joint analysis.
}
\label{Fig_GetDist}
\end{figure*}

\subsection{SDM limits}

We repeated our Fisher forecast over a range of SDM masses to determine the corresponding constraints on $\sigma_0$.
\Fig{Fig_DM_Limits} shows our forecast 95\% C.I. constraints on $\sigma_0$ 
assuming mission durations of one month, one year, and five years, corresponding to effective observation time ($t\times f_\r{eff}$) of 73h, 877h, and 4383h, respectively.
After the full mission,
Hongmeng can tighten the SDM limits by a factor of 39 compared to the current leading CMB constraints \citep{Nguyen:2021cnb},
reaching $\sigma_0 \lesssim 5.4 \times 10^{-43}\,\mathrm{cm^2}$ for $0.1\,\mathrm{MeV} \lesssim m_\chi \lesssim 0.3\,\mathrm{GeV}$.
With a more modest mission duration of one year,
these limits relax by about a factor of 2.2, 
but remain tighter by roughly 17 times than the CMB bounds.
Even with a mission duration of only one month,
Hongmeng can already achieve an improvement of a factor of four over the current CMB limits.

As pointed out in \citet{Munoz:2015bca},
at low SDM masses ($m_\chi \lesssim 1\r{GeV}$),
the SDM contribution to the IGM temperature evolution equation,
that is, \Eq{Eq_SDM_cooling_rate},
is dominated by nearly mass-independent cooling,
and thus, our $\sigma_0$ constraints are flat below $0.3\r{GeV}$.
In the opposite regime ($m_\chi \gtrsim 1$GeV),
the second term of \Eq{Eq_SDM_cooling_rate} dominates and the SDM begins to heat the IGM at a rate that roughly scales as $\rd T_\r{k}/\rd t \propto \sigma_0/m_\chi$.
Above 0.3 GeV, our constraints therefore roughly scale as $\sigma_{0} \propto m_\chi$.

\subsection{Constraints on the astrophysics}

\Fig{Fig_GetDist} shows the forecast constraints on the astrophysical parameters, 
the foreground parameters,
and $\sigma_0$ for the full Hongmeng mission assuming $m_\chi = 0.1\,\mathrm{MeV}$.
Except for $\sigma_0$, which assumes a unit of $10^{-42} \r{cm}^2$,
all parameters in the figure assume the units and parameterization (linear or $\log_{10}$) listed in \Tab{Tab_varied_params}.
Accumulation of ionizing UV radiation only causes noticeable effects on $x_\r{H}$ and consequently,
on the 21cm signal,
during the epoch of reionization (EoR),
which begins around $z \simeq 10$ in our fiducial settings.
These redshifts are not well probed by Hongmeng, which targets $11 \le z \le 46$. The green contours in \Fig{Fig_GetDist} therefore show that constraints on $f_\r{esc, 10}$ and $f_\r{esc, 7}$ are much weaker than the constraint for the SFR and X-ray parameters.

The formation of the first generation of galaxies,
that is, Pop III,
occurs much earlier than Pop II.
Therefore, 
Pop III galaxies begin to affect the 21cm signal and $T_{\rm sky}$ at much higher redshifts, 
and their influence spans a wider redshift range than that of Pop II galaxies.
Consequently,
\Fig{Fig_GetDist} shows that the prospective constraints on Pop III parameters are significantly tighter than those for Pop II.
For example,
for Pop II galaxies,
we find 1$\sigma$ errors for the X-ray strength ($\log_{10}\mathcal{L}_\r{x,II}$),
the normalization ($\log_{10} f_{\star, 10}$) and the slope ($\alpha_{\star, \r{II}}$) of the stellar fraction to be 2.2, 2.1, and 1.5,
respectively,
whereas the relevant uncertainties for the Pop III counterparts of these parameters ($\log_{10}\mathcal{L}_\r{x,III}, \log_{10} f_{\star, 7}$, and $\alpha_{\star, \r{III}}$) are 0.5, 0.1, and 0.3. This is tighter by up to a factor of 20 than for Pop II.
Hongmeng is therefore particularly suitable for constraining the properties of the first generation (Pop III) of galaxies.
In other words,
accounting for Pop III would be important for understanding the Hongmeng data.
The blue contours in \Fig{Fig_GetDist},
in which we fixed the Pop III parameters,
show that the constraints on the Pop II parameters can be significantly tightened by up to a factor of 9.
However, when the uncertainties in the properties of Pop III galaxy are not accounted for,
these limits are potentially less robust.

\subsection{Robustness of the SDM limits}
\Fig{Fig_DM_Limits} further shows the effect of Pop III galaxies on the SDM constraints,
where the dashed black line shows the expected full-mission $\sigma_0$ constraints when Pop III parameters are fixed to their fiducial values. This gives tighter (by up to a factor of 2.5) and yet potentially less robust constraints.
The effect of Pop III also depends on the parameterization of FG,
for instance,
for $N_\r{FG}=5$,
fixing Pop III parameters tightens the SDM constraints by up to a factor of 7.

\Fig{Fig_GetDist} additionally shows the correlation between $\sigma_0$ and FG parameters $p_\r{i}$.
Except for $p_3$,
which is only strongly correlated with $f_\r{esc,7}$,
$\sigma_0$ is tightly correlated with the other FG parameters. Our SDM limits can therefore be dependent on the FG parameterization.
To further examine the robustness of our results against variations in FG modeling,
we repeated our analysis for FG models with varied FG terms $N_\r{FG}$ between 0 and 10. For $N_\r{FG}=0$, we fixed $T_\r{FG}$ to the fiducial values in \Eq{Eq_TFG_data} without varying any of the FG parameters $p_i$.
Overall, we find that increasing $N_\r{FG}$ weakens the constraints, as expected.
For $N_\r{FG} \ge 7$,
the $\sigma_0$ limits converge to within 13\% (with varying Pop III parameters) and 30\% (with fixed Pop III parameters).
For the most extreme case of $N_\r{FG}=0$, 
varying Pop III parameters tighten the $\sigma_0$ limits by up to 40\%, 
while fixing them strengthens the results by more than one order of magnitude.

In our noise model, the observational uncertainty $\sigma_\r{n}$ for the global signal scales as $\sigma_\r{n} \propto t^{-1/2}$. A longer observation time $t$ therefore always lead to a higher measurement precision and thereby tightens the SDM constraints.
However,
systematic uncertainties,
such as chromatic beam effects and residuals from an imperfect FG removal,
can also contribute to the observational uncertainty \citep{2016MNRAS.455.3890M,Monsalve:2016xbk,Spinelli:2022xra},
and previous experiments such as EDGES \citep{2016MNRAS.455.3890M,Monsalve:2016xbk} and SARAS \citep{Singh:2017syr} often found a systematic noise level of $\sim$ 1mK.
While Hongmeng might be able to avoid these systematic uncertainties,
we repeated our forecast to examine their potential effects by updating the likelihood variance in \Eq{Eq_LogLike} from $\sigma^2_\r{n}$ to $\sigma^2_\r{n} + (\r{1mK})^2$,
where we used a typical irreducible noise level of 1mK.
For a mission duration of one month,
the overall noise level is dominated by the $\sigma_\r{n}$ term, and our $\sigma_0$ limits are effectively unchanged.
At longer mission durations of one and five years, 
the irreducible noise term starts to dominate,
and constraints on $\sigma_0$ can be weakened by factors of approximately 0.4 and 1,
respectively.

\subsection{Cooling-heating transition and likelihood multimodality}

\begin{figure}[t]
\begin{center}
\includegraphics[width=\textwidth/2]{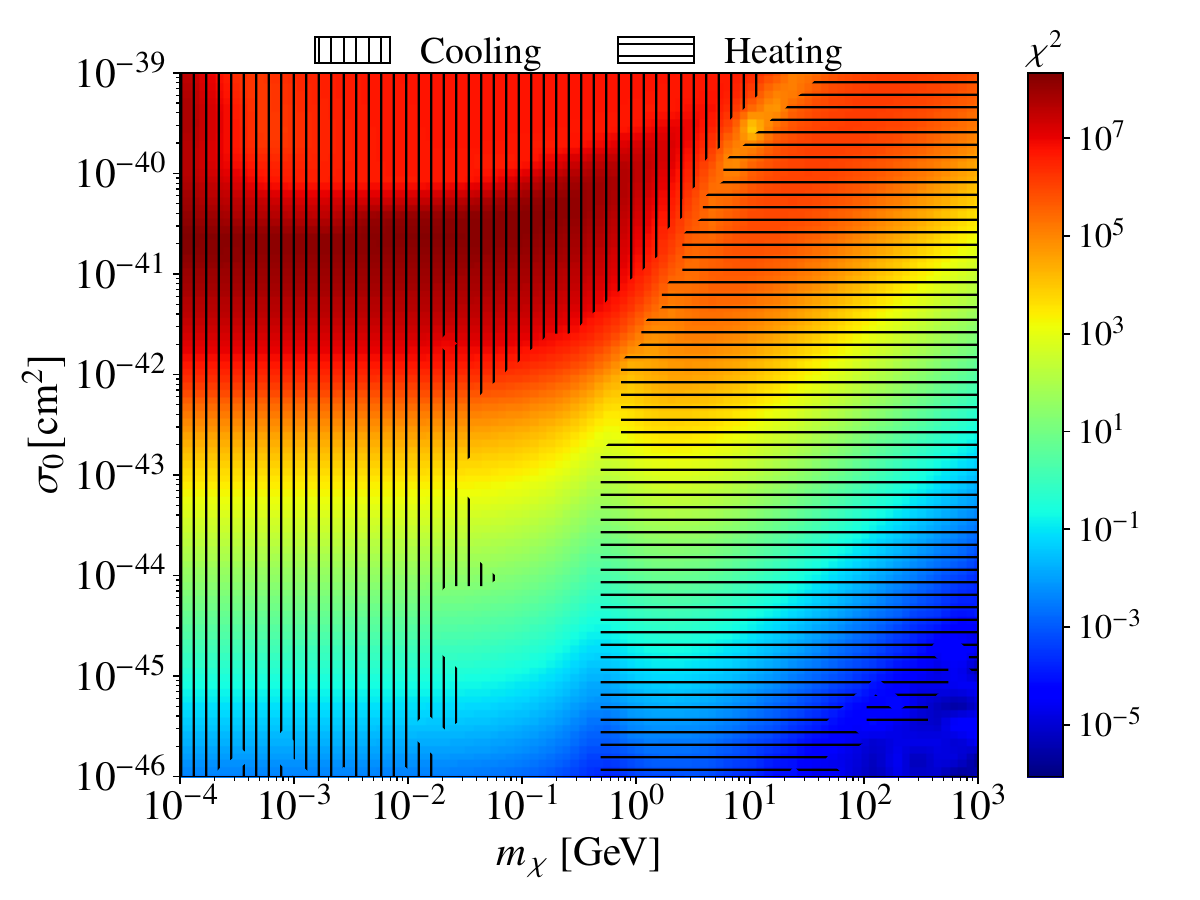}
\end{center}
\caption{
Effects of the SDM parameters on the IGM temperature and on our likelihood.
The color-coding represents the value of $\chi^2 \equiv \sum_\nu(T_\r{sky, data} - T_\r{sky, model})^2/\sigma_\r{n}^2$.
The vertical and horizontal hatches identify regions in which the SDM consistently cools or heats the IGM during $11 \lesssim z \lesssim 46$.}
\label{Fig_Cooling_VS_Heating}
\end{figure}

As we briefly discussed in \Sec{Sec_SDM},
SDM can either cool or heat IGM.
We explore this issue further in \Fig{Fig_Cooling_VS_Heating},
in which we show the distribution of $\chi^2 \equiv \sum_\nu(T_\r{sky, data} - T_\r{sky, model})^2/\sigma_\r{n}^2$ over different $\sigma_0$ and $m_\chi$ values while keeping the remaining parameters fixed to their fiducial values.
As mentioned in \citet{Munoz:2015bca},
at $m_\chi < 10\,\r{MeV}$, the SDM consistently leads to additional cooling regardless of $\sigma_0$,
and at higher masses, the SDM starts to heat the IGM at some redshifts.
Around 1 GeV,
the SDM transitions to consistently heat the IGM for $\sigma_0 \lesssim 10^{-42} \r{cm}^2$.
Around the mass scale of cooling-heating transition, 
the SDM affects the 21cm signal minimally, 
weakening the $\sigma_0$ constraints relative to the neighboring masses. 

The transition of the SDM from cooling to heating is dependent not only on the SDM mass, but also on the scattering cross section $\sigma_0$.
 \Fig{Fig_Cooling_VS_Heating} shows that
for a fixed $m_\chi$, our $\chi^2$ shows a non-monotonic behavior when $\sigma_0$ is varied,
that is,
around the cross section where the SDM transitions between cooling and heating,
the effects of the SDM on the 21cm signal are minimized and a local $\chi^2$ minimum can appear.
The likelihood can thus have complex non-Gaussian behavior and become multimodal.
We show this effect for fixed astrophysical and FG parameters, but \Eq{Eq_SDM_cooling_rate} shows that the relative weight between the heating and cooling terms also depends on the IGM temperature $T_\r{k}$, which is also dependent on the background astrophysics. The presence and location of this local $\chi^2$ minimum might therefore also be dependent on the astrophysical parameters.

This multimodality can potentially have hazardous implications for an inference analysis of real 21cm data.
Even when the true value of $\sigma_0$ is located at 0, biased inference results might be obtained, depending on the sampling strategy and
on the complexity of the background astrophysical and FG parameters.
For example, when $\sigma_0$ is sampled in log space, 
these local likelihood peaks can be underweighted and effectively disregarded when their widths are sufficiently narrow in log space.
In contrast,
sampling $\sigma_0$ in linear space can cause the same local peak to appear much broader,
and when additional parameters for the astrophysical background and FG are included,
its height can become indistinguishable from that of the true peak.
This can overweight the local maximum and might lead to a false-positive SDM detection.

Our forecasts rely on the Fisher matrix method,
which assumes that the likelihood is Gaussian in the model parameters.
As discussed above, however,
with respect to $\sigma_0$, our likelihood can be multimodal.
To examine the level of non-Gaussianity and its potential effect on our $\sigma_0$ forecast,
we used the Fisher method to approximate $\chi^2$ with respect to $\sigma_0$ as
\footnote{
In \Eq{Eq_Chi2_Fisher} we used the fact that in our fiducial setting,
$\chi^2=0$ and $\sigma_0=0$,
and the zeroth-order of $\chi^2_\r{Fisher}$ therefore vanishes and the step size of $\sigma_0$ is given simply by $\sigma_0$ itself.
},
\be
\chi^2_\r{Fisher}(\sigma_0)
=
\mathcal{F}_{\sigma_0\sigma_0}
\sigma_0^2,
\label{Eq_Chi2_Fisher}
\ee
where $\mathcal{F}_{\sigma_0\sigma_0}$ is the diagonal Fisher matrix element for $\sigma_0$.
Using the one-sigma error on $\sigma_0$ from our full-mission forecast with jointly varying Pop II and Pop III parameters as the step size
\footnote{
At small $\sigma_0$ step sizes,
$\chi^2_\r{Fisher}$ gives accurate estimates of the true $\chi^2$,
for instance,
at $\sigma_0 = 10^{-43}\r{cm^2}$,
$\chi^2_\r{Fisher}$ agrees with $\chi^2$ within 20\% for all SDM masses we explored.
},
we found that $\chi^2_\r{Fisher}$ agrees with the exact solution $\chi^2$ within 40\% for SDM mass windows $[10^{-4}, 0.3]$ GeV and $[10^{2}, 10^3]$ GeV,
indicating that the likelihood is roughly Gaussian at these masses.
However, in the $[1, 32]$ GeV mass window where the SDM transitions between cooling and heating,
$\chi^2_\r{Fisher}$ can overestimate $\chi^2$ by up to 125\%,
suggesting that beyond-linear $\chi^2$ terms become non-negligible, and our $\sigma_0$ limits are likely overly stringent.

While the Fisher formalism we used is fast and enabled us to explore a large set of astrophysical parameters,
it cannot capture likelihood multimodalities and non-Gaussianity.
Methods such as nested sampling,
Markov chain Monte Carlo (MCMC), and derivative approximation for likelihoods (DALI) \citep{Sellentin:2014zta} can better resolve these features and yield more reliable parameter constraints.
However, 
they are typically computationally prohibitive for the large set of astrophysical parameters considered here,
so we leave a full analysis to future work.

\section{Discussions}
\label{Sec_Discussions}

We explored the potential of the Hongmeng 21cm experiment of probing SDM - baryon interactions.
We jointly modeled the contributions to the observed sky temperature from the SDM,
the astrophysical background,
and the foreground.
Hongmeng is designed to operate in lunar orbit, which allows it to use the Moon as a shield from Earth RFI contaminations,
and we showed that even with a very conservative observation strategy in which the experiment operates when the Earth and Sun are both shielded by the Moon,
Hongmeng can tighten current leading SDM limits from the CMB by a factor of four after just one month of operation.
After the full five-year mission,
Hongmeng can tighten the CMB limits by a factor of 39,
reaching $\sigma_0 \lesssim 5.4\times 10^{-43} \r{cm^2}$ for SDM masses below 0.3 GeV.

In addition to the conventional Pop II galaxies,
in our astrophysical background, we also accounted for the first-generation Pop III galaxies that might be present at the high redshifts targeted by Hongmeng.
We jointly modeled the star formation,
X-ray, and UV radiation fields of the two galaxy populations to derive robust SDM constraints,
and we found that fixing the Pop III galaxy parameters
can give much tighter (by up to a factor 2.5) but less robust SDM constraints.

Measurements of the 21cm power spectrum can probe the excess cooling induced by the SDM, 
complementing global signal observations.
Although the first HERA phase II results derived from a small subset (129 hours) of the full dataset showed that the data have yet to reach the sensitivity required to detect or constrain SDM,
the full HERA phase II dataset (which has already been collected) is expected to be approximately 700 times more sensitive \citep{HERA:2025ajm},
which might be sufficient for an SDM detection or constraint.
Furthermore,
\citet{Rahimieh:2025lbf} demonstrated that with roughly 3000 hours of observation,
HERA might improve the current CMB constraints on the SDM by more than three orders of magnitude.

Depending on the SDM properties and IGM thermal states,
the SDM can either cool or heat IGM, as shown in Fig.~\ref{Fig_Cooling_VS_Heating}. This leaves a caveat that a scan across these regimes can lead to a non-Gaussian behavior in the likelihood, which is not fully captured by a Fisher analysis.
A complicated likelihood behavior can result in a multimodality that might results in a false-positive detection in an inference analysis of real 21cm data. We postpone this to future studies.

\section*{Data availability}
Data used in this work are available upon request to the corresponding author.

\begin{acknowledgements}
The authors thank Xuelei Chen, Jordan Flitter and Yuewei Wen for helpful discussions.
J.C. is thankful to NAOC for hospitality during his visit that initiated this study. 
Y.G. is partially supported by NSFC under grant no. 12275278. Y.-Z. Ma acknowledges the support from South Africa's National Research Foundation under Grants No.~150580, No.~CHN22111069370, No.~ERC250324306141.

\end{acknowledgements}

\bibliographystyle{aa}
\bibliography{aa58898-26.bib}
\end{document}